\documentclass[letterpaper, 10 pt, conference]{ieeeconf}  

\IEEEoverridecommandlockouts                              

\usepackage{graphics} 
\usepackage{epsfig} 
\usepackage{times} 
\usepackage{amsmath} 
\usepackage{amssymb}  
\usepackage{dsfont}
\usepackage{tabularx}
\usepackage{multirow}
\usepackage{booktabs} 
\usepackage[flushleft]{threeparttable}
\usepackage{subcaption}
\usepackage[labelsep=period]{caption}
\usepackage{cases}
\usepackage{xcolor}
\usepackage{colortbl}

\newtheorem{theorem}{Theorem}
\newtheorem{remark}{Remark}
\newtheorem{corollary}{Corollary}
\newtheorem{proposition}{Proposition}
\newtheorem{assumption}{Assumption}
\newtheorem{definition}{Definition}

\title{\LARGE \bf
Market Power and Withholding Behavior of Energy Storage Units}

\author{
        Yiqian Wu$^{1}$, Bolun Xu$^{2}$ and James Anderson$^{1}$
\thanks{$^{1}$Y. Wu and J. Anderson are with the Department of Electrical Engineering,
        Columbia University, New York, NY 10027, USA
        {\tt\small {\{yiqian.wu2,james.anderson\}}@columbia.edu}}%
\thanks{$^{2}$B. Xu is with the Department of Earth and Environmental Engineering, 
        Columbia University, New York, NY 10027, USA
        {\tt\small bx2177@columbia.edu}}%
}

\begin{document}

\maketitle
\thispagestyle{empty}
\pagestyle{empty}

\begin{abstract}
Electricity markets are experiencing a rapid increase in energy storage unit participation. Unlike conventional generation resources, quantifying the competitive operation and identifying if a storage unit is exercising market power is challenging, particularly in the context of  multi-interval  bidding strategies. We present a framework to differentiate strategic capacity withholding behaviors attributed to market power from inherent competitive bidding in storage unit strategies. Our framework evaluates the profitability of strategic storage unit participation, analyzing  bidding behaviors as both price takers and price makers using a self-scheduling model, and investigates how they leverage market inefficiencies. Specifically, we propose a price sensitivity model derived from the linear supply function equilibrium model to examine the price-anticipating bidding strategy, effectively capturing the influence of market power. We introduce a sufficient \textit{ex-post} analysis for market operators to identify potential exploitative behaviors by monitoring instances of withholding within the bidding profiles, ensuring market resilience and competitiveness. We discuss and verify applicability of the proposed framework to realistic settings. Our analysis substantiates commonly observed economic bidding behaviors of storage units. Furthermore, it demonstrates that significant price volatility offers considerable profit opportunities not only for participants possessing market power but also for typical strategic profit seekers. 

\end{abstract}

\section{INTRODUCTION}


Electricity markets are seeing a surging amount of battery energy storage unit participants. Jointly facilitated by the reduced cost of battery cells and removed market participation barriers~\cite{sakti2018review}, battery energy storage is becoming increasingly competitive. In California, the capacity of grid-scale battery energy storage increased by ten times in three years, from less than 500~MW in 2020 to 5,000~MW in mid-2023, and is projected to reach 10,000~MW in 2025~\cite{caiso_es}. On the other hand, participation in wholesale electricity markets to arbitrage price differences is becoming the main grid service for storage units~\cite{eia}, surpassing frequency regulation which is a specialized service that can only accommodate very limited storage capacity~\cite{ma2021data}.

Energy storage units participate in wholesale electricity markets either by self-scheduling~\cite{mohsenian2015coordinated} or submitting competitive economic bids~\cite{baker2023transferable}. In self-scheduling, storage units design their charging and discharging schedule ahead of wholesale market clearance. In economic bidding, storage units submit separate bids to charge and discharge. In both cases, storage units often perform private optimization according to their operation characteristics, degradation cost, and opportunity cost. Unlike thermal generators that directly design bid values based on their fuel cost and heat rate curve, using optimization to design self-schedule bids is critical for storage units to systematically account for factors such as future price volatility and state-of-charge constraints~\cite{chen2021pricing, krishnamurthy2017energy, bansal2022market}.

Accompanying optimized market participation is the capacity withholding of storage units. Capacity withholding occurs when a resource is purposefully limiting its supply despite the current price being higher than its real marginal production cost. Capacity withholding is often a critical sign that a participant is exercising market power by limiting the supply of a given resource in order to drive up the price and obtain higher profit. Hence this conduct is strictly monitored and regulated~\cite{baldick2005design}. However, the same regulation does not necessarily apply to energy storage units. Due to their limited energy capacity, these units must strategically target its timing to charge and discharge accounting for prices over a sequence of time, rather than focusing on the price at any single specific interval. Harvey and Hogan~\cite{harvey2001market} noted that \emph{``If a unit is energy limited, its offer price will exceed the unit's incremental cost.''}, and \emph{``This economic withholding can conceptually be distinguished from the exercise of market power.''}.

Nevertheless, as a regular participant in electricity markets, storage units can use the same withholding strategy to exercise market power as conventional generators. While storage units can conduct capacity withholding for justifiable motivations to seek  charge and discharge arbitrage opportunities, it can further increase its withholding to exercise market power. To this end, it is extremely difficult to identify whether storage unit is exercising market power. Yet, there have not been systematic studies on the intricacies of multi-interval bidding patterns, nor have they sufficiently considered the constraints of energy-limited generation resources, including energy storage systems.

We propose a systematic framework to understand how to differentiate market power from storage bids that are legitimate withholding. 
The contributions of this work are two-fold. 
\begin{enumerate}
        \item We evaluate the profitability of strategic storage unit participants in the electricity market. More precisely, we analyze their bidding behavior through a self-scheduling profit-maximization model following two sets of market roles and bidding strategies that exploit market inefficiencies: conducting competitive arbitrage as price takers or exercising market power as price makers. Uniquely, we design a price sensitivity model derived from the linear supply function equilibrium model to examine the price-anticipating bidding strategy, offering a clear view of how the exercise of market power impacts prices.
        \item From the perspective of the market operator, we provide insights into identifying market exploitation behavior accounting for the instances of withholding by analyzing the bidding profiles and the corresponding price series. This result proves to be a sufficient \emph{ex-post} tool for market efficiency monitoring. By applying our framework, we  resolve a conjecture made by Harvey and Hogan~\cite{harvey2001market} pertaining to maximized energy utilization of energy limited units during the peak (or valley for energy storage units) price periods. Additionally, we reveal the considerable profit margin exposed to both price takers and price makers attributed to high price volatility. These findings are useful for maintaining market resilience and providing robust competitiveness. 
\end{enumerate}
The main results are validated using historical price data from NYISO.

\section{MODEL FORMULATION}

\subsection{Competition and Market Power}

\begin{figure*}[t]
        \begin{subfigure}[b]{\columnwidth}
            \centering    
            \includegraphics[height=3.5cm]{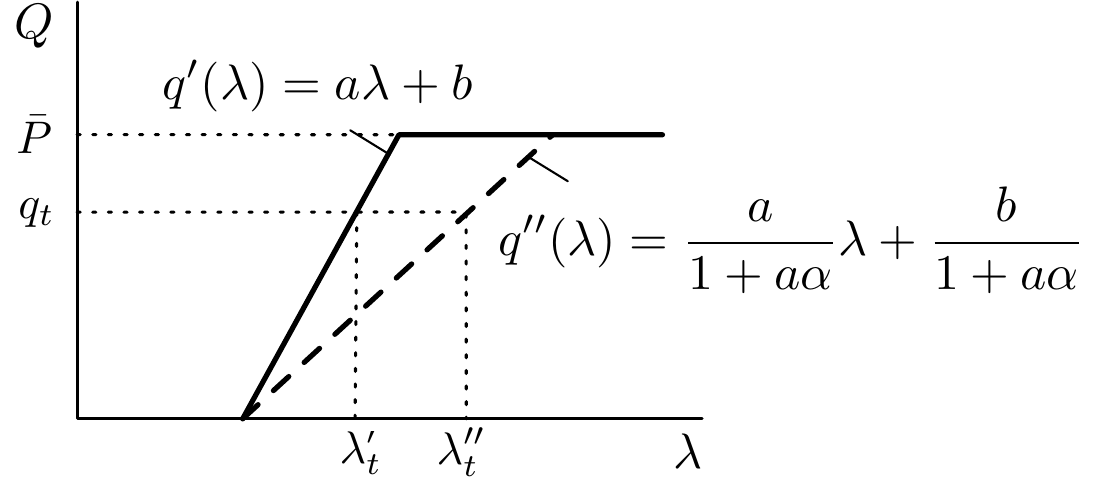}
            \caption{\small  \vspace{-0mm}}
            \label{fig:bid_curve}
        \end{subfigure}
    \hfill
         \begin{subfigure}[b]{\columnwidth}
            \centering    
            \includegraphics[height=3.5cm]{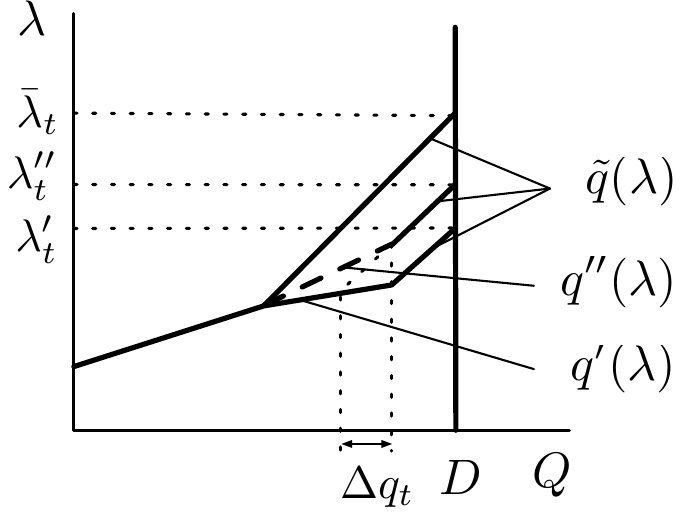}
            \caption{\small \vspace{-0mm}}
            \label{fig:withholding} 
        \end{subfigure}
        \caption{\small Bidding behavior of participants as a price taker ($q'(\lambda)$, solid line) and a price maker ($q''(\lambda)$, dashed line) and the corresponding impacts on the market outcome: (a) bid supply curves, (b) market clearing results. The axis $Q$ measures power output. The supply function $\tilde{q}(\lambda)$ represents the remainder of the aggregated supply within the system. Demand is considered inelastic at $D$, $\Delta q_t$ indicates the equivalent capacity withholding. $q'(\lambda)$ is the optimal solution to problem~\eqref{eq:bidding_taker}, and $q''(\lambda)$ is that to problem~\eqref{eq:bidding_maker}. \vspace{-1mm}}
        \label{fig:bid}
        \vspace{-1mm}
\end{figure*}

Competitive arbitrage and the exercise of market power are both mechanisms that leverage market inefficiencies. 
Arbitrage involves market participants leveraging price difference and the ability to leverage this at no risk. In contrast,
the exercise of market power involves market participants deliberately reducing their production below competitive levels in order to increase market clearing prices, thereby enhancing their profitability~\cite{harvey2001market}. 
In this work, we aim to analyze the consequences of these bidding strategies by modeling the optimal strategy making process for multi-interval bidding  from the perspective of a single storage unit.
The underlying theory of market power is that the market clearing price will be influenced by the offers or bids from certain participants. 
We explicitly model such impact on the price  at time $t$ using a non-negative price sensitivity parameter~$\alpha_t$. The price at time $t$ is modeled  as
\begin{align}
  \lambda_t = \bar{\lambda}_t - \alpha_t q_t \label{eq:price_sensitivity}
\end{align}
where $\lambda_t$ is the influenced price, $\bar{\lambda}_t$ is the nominal clearing price and $q_t$ is the dispatch decision. 
It is important to clarify that $\bar{\lambda}_t$ does not represent the competitive price; rather, it signifies the price in the absence of the participation of a specific market participant.\footnote{A similar price sensitivity model is introduced in~\cite{reinisch2006}. However, the model in~\cite{reinisch2006} quantifies price impact via the amount of withheld production, $\lambda_t = \tilde{\lambda}_t + \alpha_t \Delta q_t$. This approach does not capture the immediate price drop following the entry of a price maker. Additionally, it calculates withheld production based on the competitive level, which complicates the analysis.}

Equation~\eqref{eq:price_sensitivity} captures the fact that, as a market participant possessing market power, their participation in the market ($q_t \neq 0$) naturally brings down the market price ($\lambda_t < \bar{\lambda}_t$).
This effect is particularly notable when the participant serves as a large-scale pivotal supplier, whose power capacity is essential for easing potential power shortages in the system. Hence, the nominal price $\bar{\lambda}_t$ functions as an indicator of the system's load level or supply capacity. 
Additionally, offering more power to the market lowers the price. Therefore,  practices such as physical capacity withholding ($q_t < \bar{P}$, where $\bar{P}$ denotes the power capacity) can raise market prices, generating a profit margin for the withholding entity and possibly others but also diminishing social welfare by affecting the overall market equilibrium.


To understand the bidding behavior of participants anticipating their impact on prices and the consequent market outcomes, we consider a bid-based market that clears by relying on affine supply functions~\cite{baldick2004}:
\begin{align}
        q_t(\lambda_t) = a_t \lambda_t + b_t \label{eq:bidding_curve}
\end{align}
where the parameter pair ($a_t, b_t$) characterizes the offer chosen by the participant for time interval~$t$.

Generally, a rational market participant designs their bid supply function $q_t(\lambda_t)$ by solving a profit-maximization problem using price forecast $\hat{\lambda}_t$  over a future time period~$t\in~\mathcal{T}$: %
\begin{align}
        q_t(\lambda_t) = \underset{a_t,b_t}{\mathrm{argmax}}\ \pi(q_t(\lambda_t);\hat{\lambda}_t)  \quad\mathrm{s.t.} \ \eqref{eq:bidding_curve}
         \label{eq:bidding_taker}
\end{align}
where $\pi = \sum_{t \in \mathcal{T}} \hat{\lambda}_t q_t(\lambda_t) - C\left(q_t(\lambda_t)\right)$ represents the total profit and $C(\cdot)$ is a given strictly convex cost function.

Ideally, the supply function $q_t(\lambda_t)$ should reflect the actual operation costs of production in a competitive market. However, bids that incorporate price sensitivity, as determined by solving
\begin{align}
        q_t(\lambda_t) = \underset{a_t,b_t}{\mathrm{argmax}}\ \pi(q_t(\lambda_t),\bar{\lambda}_t - \alpha  q_t(\lambda_t))  \quad\mathrm{s.t.} \ \eqref{eq:bidding_curve}
        \label{eq:bidding_maker}
\end{align}
may obscure the true cost information.
Assuming a quadratic cost function $C(\cdot)$, we  illustrate the variation in bidding behavior and the consequent market outcome  in Fig.~\ref{fig:bid} (for simplicity, the subscript $t$ in the bidding functions is neglected in the figures). $q'(\lambda)$ is the optimal solution to problem~\eqref{eq:bidding_taker}, and $q''(\lambda)$ is that to problem~\eqref{eq:bidding_maker}. As shown in Fig.~\ref{fig:bid}(\subref{fig:bid_curve}), the bidding curve shifts from $q'(\lambda)$ to $q''(\lambda)$ due to the consideration of price sensitivity. This shift results in the market clearing price increasing from $\lambda'_t$ to $\lambda''_t$, while maintaining the same level of power output $q_t$, indicating a clear reduction in economic efficiency. This alteration in the bidding curve can be seen as cost-related withholding, i.e., economic withholding.
The market clearing outcome related to this behavior is shown in Fig.~\ref{fig:bid}(\subref{fig:withholding}). Here, we use $\tilde{q}(\lambda)$ to denote the remainder of the aggregated supply function within the system. It can be seen that, as two primary approaches of conducting capacity withholding or further exercising market power~\cite{harvey2001market}, economic withholding through the bid supply function is equivalent to physical withholding of the amount~$\Delta q_t$ from the perspective of market outcomes. For simplicity, we will focus on the effects of withholding practices and refer to this as capacity withholding throughout the remainder of the paper, without loss of generality.

Additionally, it can be inferred from Fig.~\ref{fig:bid}(\subref{fig:withholding}) that the determination of the price sensitivity parameter~$\alpha_t$ is influenced by the slopes of the system supply function ($\tilde{q}(\lambda)$ in the figure), the demand function ($D$ in the figure), and the unit supply function.
Although delving into the specifics of this process extends beyond the scope of this work, it highlights the necessity for market participants with market power to obtain the knowledge of the system demand level and the bids from other participants to accurately estimate their influence on market outcomes and making corresponding bidding decisions. {Interested readers may refer to~\cite{reinisch2006} for more details.}



We define market participants based on their bidding strategies and their impacts on the market clearing prices as follows:
\begin{definition}[Price Taker]
        A market participant is a price taker if it accepts the existing prices as given and lacks the market share to influence market prices on its own.
\end{definition}
\begin{definition}[Price Maker]
        A market participant is a price maker if they anticipate the influence of their bids on market prices. A price maker has sufficient knowledge of the system status, such as demand levels and the bids of other participants.
\end{definition}

A more thorough analysis of the bidding strategies associated with these two categories of market participants will be discussed in later sections.

\begin{remark}
        Our price sensitivity model can be generalized to scenarios where the bidding entity operates with other forms of cost functions, or participates in the market with constant cost bids or quantity bids.
\end{remark}


\subsection{Strategic Bidding of Energy Storage Units}
\allowdisplaybreaks
We adopt a convex self-scheduling model to characterize the bidding strategies of energy storage units, utilizing a series of price point forecasts $\hat{\lambda}_t$ for all $  t \in \mathcal{T}$, where $\mathcal T = \{1,2,\hdots, T\}$~\cite{xu2020}. The profit from future intervals can be regarded as the opportunity value at the current time interval. Most existing market designs require storage unit operators to make dispatch decisions based on future price estimates. This strategy  reflects  the allocation of limited output into a profile that maximizes total profit over the scheduling period. {The solution to the profit-maximization problem forms the control policy or decision of the storage units.} For ease of exposition, we assume that the storage unit accurately estimates their impact on market prices with $\alpha_t$:
\begin{subequations}
\label{eq:sses}
\begin{align}
       \underset{p_t,b_t, e_t}{\text{maximize}}  &\quad \sum_{t \in \mathcal{T}}\hat{\lambda}_t(p_t-b_t)  \label{eq:storage_profit}\\
\mathrm{s.t.}
&\quad 0\le p_t,b_t \le \bar{P},\, \quad \forall t \in \mathcal{T}  \label{eq:pch_limit} \\
& \quad p_t = 0 \textrm{ if } \hat{\lambda}_t <0,\, \quad \forall t \in \mathcal{T}\label{eq:storage_idle}\\
&\quad e_t - e_{t-1} = -\frac{p_t}{\eta} + b_t \eta,\,\quad  \forall t \in \mathcal{T} \label{eq:soc_dynamic}\\
& \quad 0\le e_t \le E,\, \quad \forall t \in \mathcal{T} \label{eq:soc_bar}
\end{align}
\end{subequations}
where $p_t$ and $b_t$ denote the amount of energy discharged and charged  respectively over time interval $t$. We refer to intervals when $0<p_t<\bar{P}$ or $0<b_t<\bar{P}$ for any $t\in \mathcal{T}$ as \emph{withholding intervals}. Objective~\eqref{eq:storage_profit} represents the total profit $\pi$ over the future period $t\in \mathcal{T}$. For price makers, it follows from~\eqref{eq:price_sensitivity} that the price forecast takes the form based on the price sensitivity: $\hat{\lambda}_t = \bar{\lambda}_t - \alpha_t (p_t-b_t)$. {Note that incorporating both forms of the price for profit calculation in~\eqref{eq:storage_profit} preserves the convexity of model~\eqref{eq:sses}.} Constraint~\eqref{eq:pch_limit} captures the charging/discharging power lower and upper bounds.
Constraint~\eqref{eq:storage_idle} ensures that the storage unit does not  discharge during periods of negative pricing, a sufficient condition to preclude  simultaneous charging and discharging~\cite{xu2020}.
The inter-temporal relationship of the state of charge (SoC) $e_t$ is defined in~\eqref{eq:soc_dynamic} with charging and discharging efficiency parameter $\eta \in (0,1]$. Inequality~\eqref{eq:soc_bar} models energy storage capacity. {We define storage units as operating in \emph{idle scenarios} if they remain on standby according to model~\eqref{eq:sses} throughout the scheduling period~$\mathcal{T}$, i.e., $p_t = 0,~b_t = 0$, for all $t \in \mathcal{T}$.}

\begin{remark}
        Although the  framework for energy storage unit bidding in this paper is based on self-scheduling and {neglects the operating cost $C(\cdot)$ for profit calculation}, it can be easily adapted to allow economic bidding~\cite{mohsenian2015coordinated}, accounting for the costs associated with storage discharge, such as those related to unit degradation, operation and maintenance, in its profit calculation. The  results discussed in this paper would still apply in this extended context. 
\end{remark}

Note that  our strategic bidding model is deterministic. We have intentionally ignored stochastic effects in order to simplify our analysis. Introducing uncertainty would necessitate  more sophisticated bidding strategies which would unnecessarily complicate the analysis and obfuscate the underlying mechanisms at play. Consequently, we plan to explore these more realistic settings in  future work.

\section{MAIN RESULTS}
We now present our framework for energy storage bidding, demonstrating the distinction between a storage unit exercising  market power and competitive capacity withholding. In the sequel, we  examine the bidding strategies of both \emph{price takers} and \emph{price makers}.

{
\begin{theorem}
        Assume the energy storage unit behaves rationally and designs its bid by solving the profit-maximization problem~\eqref{eq:sses} using price forecast $\hat{\lambda}_t$  over a horizon of length $N*T$ for $N$ bidding scheduling periods, with each period of length $T$. Given a series of observed storage power output profiles $\{p_t,~b_t\}$ and market clearing prices $\{{\lambda}_t\}$ for all $t \in \tilde{\mathcal{T}}$, where $\tilde{\mathcal{T}}=\{1,2,\ldots, NT\}$, the storage unit is not evidently exercising market power, if the following conditions are satisfied:
        \begin{enumerate}
                \item The number of withholding intervals is no larger than the number of non-idle scheduling periods $N'$, i.e., $\sum_{t\in \tilde{\mathcal{T}}}\mathds{1}_{\{0<p_t<\bar{P}\}}+ \sum_{t\in\tilde{ \mathcal{T}}}\mathds{1}_{\{0<b_t<\bar{P}\}}\le N' \le N$, 
                \item The price-decision relationship is consistent with Proposition~\ref{prop:taker_price}.
        \end{enumerate}
        \label{theorem:oneinterval}
\end{theorem}
}

{Proposition~\ref{prop:taker_price} reveals the relationship between the control decisions of a price taker and the prevailing market prices. Details of Proposition~\ref{prop:taker_price} will be provided in subsequent subsections.}

Theorem~\ref{theorem:oneinterval} suggests that the market operator can conduct \textit{ex-post} analysis according to the observed storage power output profiles and market clearing price series to examine the exercise of market power with the aim of monitoring market efficiency. This method aligns with the \emph{``after-the-fact standard''} approach for analysis as recommended by Harvey and Hogan in~\cite{harvey2001market}. Note that from the market operator's perspective, that actual competitive bids and the resulting clearing prices are not accessible. The available information for conducting market efficiency analysis comprises solely the effective storage power output profiles and the corresponding prices. The presence of partial operation intervals during the observation period indicates instances of storage  unit conducting capacity withholding. {Merely counting these instances below a predefined threshold and examining the prices associated with these instances} offers a practical and efficient method to differentiate the causes of capacity withholding -- whether or not it results from market power exercise or competitive behavior. 

The proof of Theorem~\ref{theorem:oneinterval} will compromise an examination of bidding strategies derived from simplified individual profit maximization problems discussed in subsequent subsections. Strictly speaking, we should directly deal with the bidding model~\eqref{eq:sses}; however, we will initially introduce a set of assumptions to facilitate preliminary discussions and propositions. These assumptions will later be revisited and removed, allowing us to generalize the conditions and validate the applicability of Theorem~\ref{theorem:oneinterval}.


In the simplified profit maximization model, we focus on scenarios with short clearing periods, e.g., $T=2,3$. Given the short duration of storage charging and discharging, we assume the storage capacity constraint~\eqref{eq:soc_bar} is strictly satisfied throughout the scheduling period~$\mathcal{T}$ and neglect it in the model formulation:
\begin{assumption}[Unsaturated SoC]
        During the clearing period, the energy storage SoC remains strictly feasible, i.e., $ 0< e_t < E$ for all $t \in \mathcal{T}$.
        \label{ass:short_period}
\end{assumption}

 In light of Assumption~\ref{ass:short_period}, the SoC evolution constraint~\eqref{eq:soc_dynamic}   reduces to:
\begin{align}
        \sum_{t \in \mathcal{T}}\frac{p_t}{\eta}-b_t\eta = 0.
\end{align}

\begin{assumption}[Positive profit operation]
        The storage unit remains idle, i.e., $p_t = 0,~b_t = 0$, for all $t \in \mathcal{T}$, if the induced profit is zero, i.e., $\pi=0$.
        \label{ass:storage_idle}
\end{assumption}

\begin{assumption}[Non-negative pricing]
        Prices remain non-negative, $\lambda_t\ge0$, for all $t \in \mathcal{T}$.
        \label{ass:nonneg_price}
\end{assumption}

Based on Assumption~\ref{ass:nonneg_price}, we omit the constraint~\eqref{eq:storage_idle} in the bidding model,  however the non-simultaneous charging and discharging condition \emph{does} still hold.


\subsection{Withholding as Price Taker}
First, we aim to analyze the withholding bidding behavior assuming that the energy storage unit participates in the market as a price taker. From Assumptions~\ref{ass:short_period}-\ref{ass:nonneg_price}, the 
 corresponding bidding model is:
\begin{subequations}
  \label{eq:sses3taker}
\begin{align}
  \underset{p_t,b_t}{\text{maximize}}&\quad \sum_{t \in \mathcal{T}}\hat{\lambda}_t (p_t-b_t)  \label{eq:obj_profit_taker}\\
  \mathrm{s.t.}& \quad \sum_{t \in \mathcal{T}}\frac{p_t}{\eta}-b_t\eta = 0 :\, \theta \label{eq:power_bal}\\
  &\quad 0\le p_t \le \bar{P}  :\, \delta_t^-,\delta_t^+,\, \forall t \in \mathcal{T} \label{eq:p_bar}\\
  &\quad 0\le b_t \le \bar{P}  :\, \beta_t^-,\beta_t^+,\, \forall t \in \mathcal{T} \label{eq:b_bar}
\end{align}
\end{subequations}
where the corresponding dual variable is defined after each constraint.
Note that although we formulate the bidding model of price takers using $\hat{\lambda}_t$ in the objective~\eqref{eq:obj_profit_taker}, in the presence of price makers,~$\hat{\lambda}_t$ will be replaced by the influenced price based on ~\eqref{eq:price_sensitivity} to determine the ultimate dispatch decision and profit.

The  Lagrangian function of~\eqref{eq:sses3taker} is:
\begin{align}
        L = &\sum_{t \in \mathcal{T}}\Big(\hat{\lambda}_t (p_t-b_t) + \theta(\frac{p_t}{\eta}-b_t\eta)\\ \nonumber
        & + \delta_t^- p_t - \delta_t^+(p_t-\bar{P})+ \beta_t^- b_t - \beta_t^+(b_t-\bar{P})\Big),
\end{align}
and the corresponding KKT conditions are ($\forall t \in \mathcal{T}$):
\begin{subequations}
  \begin{align}
    & \frac{\partial L}{\partial p_{t}} = \hat{\lambda}_{t} +\frac{\theta}{\eta}+ \delta_{t}^- - \delta_{t}^+ = 0 \label{eq:kkt_p}\\
    & \frac{\partial L}{\partial b_{t}} = -\hat{\lambda}_{t} -\theta\eta+ \beta_{t}^- - \beta_{t}^+ = 0 \label{eq:kkt_b}\\
    & \sum_{t \in \mathcal{T}}\frac{p_t}{\eta}-b_t\eta = 0 \label{eq:kkt_bal}\\
    & 0\le p_{t}, ~ b_{t}  \le \bar{P} \label{eq:kkt_primal}\\
    & \delta_{t}^-, ~ \delta_{t}^+,~ \beta_{t}^-, ~\beta_{t}^+~ \ge 0 \label{eq:kkt_dual}\\
    & \delta_{t}^- p_{t} = 0, \, \delta_{t}^+ (p_{t} - \bar{P} ) = 0  \label{eq:kkt_com1}\\
    & \beta_{t}^- b_{t} = 0, \,\beta_{t}^+ (b_{t} - \bar{P} ) = 0 \label{eq:kkt_com2}
  \end{align}
  \label{eq:kkt_taker}
\end{subequations}
where Eqs.~\eqref{eq:kkt_p} and \eqref{eq:kkt_b} correspond to stationarity conditions, constraints~\eqref{eq:kkt_bal} and \eqref{eq:kkt_primal} to primal feasibility, constraint~\eqref{eq:kkt_dual} to dual feasibility, Eqs.~\eqref{eq:kkt_com1} and \eqref{eq:kkt_com2} to complementary slackness. Given that the optimization model~\eqref{eq:sses3taker} is a linear program, there is zero duality gap, yielding KKT conditions both necessary and sufficient for optimality.

\begin{proposition}
        For a strategic price taker making bidding decisions based on model~\eqref{eq:sses3taker}, the bidding decisions $\{p^*_t,~b^*_t \}$ throughout the period $\mathcal{T}$ will include at least one interval $t$ at capacity, i.e.,  $p^*_t = \bar{P}$ or $b^*_t = \bar{P}$, except for the idle scenarios. \label{prop:full}
\end{proposition}
\begin{proof} 
         Suppose the unit withholds a certain amount of capacity throughout the period $\mathcal{T}$: $0< p^*_t,~b^*_t < \bar{P},\, \forall t \in \mathcal{T}' \subseteq \mathcal{T}$, where $\mathcal{T}'$ is the set of non-idle intervals. According to the complementary constraints~\eqref{eq:kkt_com1} and \eqref{eq:kkt_com2}, we have ${\delta_t^{+}}^*,~{\beta_t^{+}}^*=0,\, \forall t \in \mathcal{T}'$. The non-idle condition suggests that there exists $t_i,t_j \in \mathcal{T}'$, that  $p_{t_i}^* > 0$ and $b_{t_j}^*>0$, thus we have ${\delta_{t_i}^-}^*|_{p_{t_i}^*>0} = 0$ and $ {\beta_{t_j}^-}^*|_{b_{t_j}^*> 0} =0 $  based on~\eqref{eq:kkt_com1} and \eqref{eq:kkt_com2}, then the corresponding stationary conditions~\eqref{eq:kkt_p},~\eqref{eq:kkt_b} for these intervals are 
        \begin{align}
        \left\{
            \begin{aligned}
              &\hat{\lambda}_{t_i}+\frac{\theta}{\eta}  =  0, \forall t_i \in \{t_i~|~p_{t_i}^*>0, t_i \in \mathcal{T}'\},\\
              &-\hat{\lambda}_{t_j}-\theta\eta  =  0, \forall t_j \in  \{t_j~|~b_{t_j}^*>0, t_j \in \mathcal{T}'\}.
            \end{aligned}
            \right. 
            \label{eq:ex}
        \end{align}
        Clearly, solutions to~\eqref{eq:ex} need to satisfy $\hat{\lambda}_{t_i}\eta^2 = \hat{\lambda}_{t_j}$. Given the profit is calculated as~\eqref{eq:obj_profit_taker}, where $p_t$ and $b_t$ satisfy~\eqref{eq:power_bal}, prices satisfying $\hat{\lambda}_{t_i}\eta^2 = \hat{\lambda}_{t_j}$ result in a profit $\pi=0$ . Based on Assumption~\ref{ass:storage_idle}, the considered intervals keep idle, which violates the non-idle condition. 
\end{proof}

\begin{proposition}
For a strategic price taker making bidding decisions based on model~\eqref{eq:sses3taker}, assuming the storage unit is not idle throughout the scheduling period $\mathcal{T}$, then 
\begin{enumerate}
        \item there must exist $\hat{\lambda}_{\mathrm{min}} <~\hat{\lambda}_{\mathrm{max}}$, where $\hat{\lambda}_{\mathrm{min}} = \mathrm{min}_{t\in \mathcal{T}}\{\hat{\lambda}_t\}$ and $\hat{\lambda}_{\mathrm{max}} = \mathrm{max}_{t\in \mathcal{T}}\{\hat{\lambda}_t\}$, \label{prop:2_1}
        \item the storage unit operates at capacity either at the highest price interval ${t_{\hat{\lambda}}}_{\mathrm{max}}$ or the lowest price interval ${t_{\hat{\lambda}}}_{\mathrm{min}}$:
        ${p^*_{t_{\hat{\lambda}}}}_{\mathrm{max}} = \bar{P}$ or ${b^*_{t_{\hat{\lambda}}}}_{\mathrm{min}} = \bar{P}$. \label{prop:2_2}
\end{enumerate}

        \label{prop:full_price}
\end{proposition}

\begin{proof}
    We first prove~\ref{prop:2_1}) that the prices throughout the scheduling period $\mathcal{T}$ are not identical. This can be done by assuming that all prices are identical: $\hat{\lambda} = \hat{\lambda}_{1} = \hat{\lambda}_{2}=  \ldots = \hat{\lambda}_{t_T}$. Thus the profit is calculated as $\pi = \sum_{t \in \mathcal{T}} \hat{\lambda}_t(p^*_t-b^*_t) = \sum_{t \in \mathcal{T}} \hat{\lambda}(p^*_t-b^*_t)$. Given the storage unit charge and discharge reset constraint~\eqref{eq:power_bal}, we have $\sum_{t \in \mathcal{T}}p^*_t = \sum_{t \in \mathcal{T}}b^*_t \eta^2$, which yields a negative profit $\pi$, violating the non-idle condition. Thus, there must exist distinct minimum and maximum prices $\hat{\lambda}_{\mathrm{min}}$ and $ \hat{\lambda}_{\mathrm{max}}$.

Next, we prove the storage unit operates at capacity either at the highest or the lowest price interval as~\ref{prop:2_2}). The counter conditions include i) both of these two intervals are at partial capacity, ii) one of these two intervals is at partial capacity and another is idle, or iii) both of these two intervals are idle. For case i), given ${p^*_{t_{\hat{\lambda}}}}_{\mathrm{max}},~ {b^*_{t_{\hat{\lambda}}}}_{\mathrm{min}}~\in (0,\bar{P})$, we have ${{\delta^{-}}^*_{t_{\hat{\lambda}}}}_{\mathrm{max}},~{{\delta^{+}}^*_{t_{\hat{\lambda}}}}_{\mathrm{max}},~{{\beta^{-}}^*_{t_{\hat{\lambda}}}}_{\mathrm{min}},~{{\beta^{+}}^*_{t_{\hat{\lambda}}}}_{\mathrm{min}}=0$ based on~\eqref{eq:kkt_com1} and~\eqref{eq:kkt_com2}. Following Proposition~\ref{prop:full}, we assume the units operates at capacity at interval $t_0$. Thus we have for these intervals:
\begin{align}
  &\mathrm{if}~{p^*_{t_0}} = \bar{P},~\mathrm{then}\left\{
    \begin{aligned}
      &\hat{\lambda}_{\mathrm{max}}+\frac{\theta}{\eta} =  0,\\
      &-\hat{\lambda}_{\mathrm{min}}-\theta\eta   =  0,\\
      & \hat{\lambda}_{t_0}+\frac{\theta}{\eta} -\delta_{t_0}^+ =  0,
    \end{aligned}
  \right. 
  \label{eq:prop2_1}
\\
\mathrm{or}\nonumber\\ 
      & \mathrm{if}~{b^*_{t_0}} = \bar{P},~\mathrm{then}\left\{
        \begin{aligned}
          &\hat{\lambda}_{\mathrm{max}}+\frac{\theta}{\eta}  =  0,\\
          &-\hat{\lambda}_{\mathrm{min}}-\theta\eta   =  0,\\
          & -\hat{\lambda}_{t_0}-\theta\eta-\beta_{t_0}^+  =  0 .
        \end{aligned}
      \right. 
      \label{eq:prop2_2}
  \end{align}

Solutions to~\eqref{eq:prop2_1} need to satisfy $\hat{\lambda}_{\mathrm{max}} = \hat{\lambda}_{t_0}- \delta_{t_0}^+$. Given non-negative $\delta_{t_0}^+$, such solutions don't exist; similar for solutions to~\eqref{eq:prop2_2}. Therefore, case i) doesn't exist. Similar proof can be provided for case ii) and case iii). Therefore, the storage unit must be at capacity either at ${t_{\hat{\lambda}}}_{\mathrm{max}}$ or~${t_{\hat{\lambda}}}_{\mathrm{min}}$.
\end{proof}

\begin{remark}
    Propositions~\ref{prop:full} and \ref{prop:full_price} suggest that during the scheduling period, the storage unit will operate at its capacity for at least one interval, either at the highest or the lowest price point. This insight is readily applicable to scenarios where the SoC of the storage unit is constrained throughout the period. In situations where the storage unit's maximum output is limited by its capacity, it can still be asserted, without loss of generality, that the storage unit will maximize its energy utilization during the period of highest or lowest prices. This principle aligns with the conjecture made in Section III.A of~\cite{harvey2001market}, which stated that \emph{``for energy limited unit, efficient pricing would fully utilize the energy of the unit in the highest price hours over the period of the limitation''}. Our findings validate this conjecture and also extend it by exploring its implications for the bidding strategies of energy storage units.  
\end{remark}

\begin{proposition}
        For a strategic price taker making bidding decisions based on model~\eqref{eq:sses3taker}, given strictly heterogeneous prices $\hat{\lambda}_{1} \neq \hat{\lambda}_{2} \ldots \neq \hat{\lambda}_{T}$, the bidding decisions $\{p^*_t,~b^*_t \}$ throughout the period $\mathcal{T}$ should include one and only one interval $t$ at partial capacity, $0< p^*_t < \bar{P}$ or $0< b^*_t < \bar{P}$, except for the idle scenarios.
        \label{prop:partial}
\end{proposition}
\begin{proof} 
To prove that there exists one and only one partial interval throughout the bidding period, we first prove there is at least one interval with partial capacity and then prove at most one interval with partial capacity. 

First, we prove at least one interval operates at partial capacity. For non-idle intervals $\mathcal{T}' \subseteq \mathcal{T}$, suppose the unit operates at full capacity $p^*_t = \bar{P}$ or $b^*_t = \bar{P}, \, \forall t \in \mathcal{T}'$. Obviously, it violates the charging and discharging balance $\sum_{t\in \mathcal{T}'} \frac{p_t^*}{\eta}-b_t^*\eta=0$.

Second, we prove at most one interval operates at partial capacity. Suppose there are more than one interval that operate at partial capacity. 
For partial intervals, we have ${\delta_t^+}^*,~{\beta_t^+}^*=0$ and ${\delta_t^-}^*=0$ or ${\beta_t^-}^*=0, \, \forall t \in \{t|0< p^*_{t} <\bar{P}\ ||\  0< b^*_{t} <\bar{P}, t\in \mathcal{T}\}$ based on~\eqref{eq:kkt_com1} and \eqref{eq:kkt_com2}, thus
  \begin{equation}
      \left\{
        \begin{aligned}
          &\hat{\lambda}_t+\frac{\theta}{\eta} +\delta_t^- =  0,\\
          &-\hat{\lambda}_t-\theta\eta  =  0,
        \end{aligned}
      \right.
      \quad \text{or} \quad
      \left\{
        \begin{aligned}
          &\hat{\lambda}_t+\frac{\theta}{\eta} =  0,\\
          &-\hat{\lambda}_t-\theta\eta  +\beta_t^- =  0,
        \end{aligned}
      \right.  \nonumber
  \end{equation}
  leading to 
  \begin{equation}
      \left\{
        \begin{aligned}
          &\hat{\lambda}_{t_1}+\frac{\theta}{\eta}  =  0,\\
          &-\hat{\lambda}_{t_2}-\theta\eta  =  0,
        \end{aligned}
      \right.
      \ \text{or} \
      \left\{
        \begin{aligned}
          &\hat{\lambda}_{t_1}+\frac{\theta}{\eta}  =  0,\\
          &\hat{\lambda}_{t_2}+\frac{\theta}{\eta}  =  0,\\
        \end{aligned}
      \right.
      \ \text{or} \
      \left\{
        \begin{aligned}
          &-\hat{\lambda}_{t_1}-\theta\eta  =  0,\\
          &-\hat{\lambda}_{t_2}-\theta\eta  =  0.
        \end{aligned}
      \right. 
      \label{eq:partial_pair}
  \end{equation}
  All pairs in~\eqref{eq:partial_pair} would have led to idle intervals as discussed in the proof for Proposition~\ref{prop:full}, or require equal prices, which violate either the non-idle or the strictly heterogeneous price assumption.         
\end{proof}

{In competitive markets, i.e., market participants bid as price takers according to model~\eqref{eq:sses3taker}, we have the following price-decision relationship:

\begin{proposition}
        Given a series of prices $\hat{\lambda}_{t}$  throughout the period $\mathcal{T}$, a strategic price taker makes bidding decisions $\{p^*_t,~b^*_t \}$ based on model~\eqref{eq:sses3taker}. Denote the set of discharge withholding intervals $\{u \in \mathcal{T}|\mathds{1}_{\{0<p_u<\bar{P}\}} =1\}$ and charge withholding intervals  $\{v \in \mathcal{T}|\mathds{1}_{\{0<b_v<\bar{P}\}} =1\}$, then  the bidding decisions  satisfy:
        \begin{enumerate}
                \item if the unit discharges at capacity during interval $x$, i.e., $p^*_x=\bar{P}$, then $\hat{\lambda}_x>\hat{\lambda}_u$ and $\hat{\lambda}_x>\frac{ \hat{\lambda}_v}{\eta^2}$ ,
                \item if the unit charges at capacity during interval $y$, i.e., $b^*_y=\bar{P}$, then $\hat{\lambda}_u > \frac{ \hat{\lambda}_y}{\eta^2}$ and $\hat{\lambda}_v>\hat{\lambda}_y$,
                \item if the unit keeps idle during interval $z$, i.e., $p^*_z = b^*_z=0$, then $\frac{ \hat{\lambda}_z}{\eta^2}>\hat{\lambda}_u>\hat{\lambda}_z$ and $ \hat{\lambda}_z>\hat{\lambda}_v>\hat{\lambda}_z\eta^2$.
        \end{enumerate} 
        \label{prop:taker_price}
\end{proposition}}
\begin{proof}
        Following the KKT conditions in~\eqref{eq:kkt_taker} and similar to the previous discussion, we have:
        
        \begin{subequations}
        \noindent for a discharge withholding interval $u$
        \begin{equation}
                \left\{
                  \begin{aligned}
                    &\hat{\lambda}_u+\frac{\theta}{\eta} =  0,\\
                    &-\hat{\lambda}_u-\theta\eta  +\beta_u^- =  0,
                  \end{aligned}
                \right.   \nonumber
            \end{equation}
        for a charge withholding interval $v$
        \begin{equation}
                \left\{
                  \begin{aligned}
                    &\hat{\lambda}_v+\frac{\theta}{\eta} +\delta_v^- =  0,\\
                    &-\hat{\lambda}_v-\theta\eta  =  0,
                  \end{aligned}
                \right. \nonumber
        \end{equation}
        for a discharge at-capacity interval $x$
        \begin{equation}
                \left\{
                  \begin{aligned}
                    &\hat{\lambda}_x+\frac{\theta}{\eta} -\delta^+_x =  0,\\
                    &-\hat{\lambda}_x-\theta\eta  +\beta_x^- =  0,
                  \end{aligned}
                \right. \nonumber
            \end{equation}
        for a charge at-capacity interval $y$
        \begin{equation}
                \left\{
                  \begin{aligned}
                    &\hat{\lambda}_y+\frac{\theta}{\eta} +\delta^-_y =  0,\\
                    &-\hat{\lambda}_y-\theta\eta  -\beta_y^ + =  0,
                  \end{aligned}
                \right.  \nonumber
            \end{equation}
        and for an idle interval $z$
        \begin{equation}
                \left\{
                  \begin{aligned}
                    &\hat{\lambda}_z+\frac{\theta}{\eta} +\delta^-_z =  0,\\
                    &-\hat{\lambda}_z-\theta\eta  +\beta_z^- =  0.
                  \end{aligned}
                \right.  \nonumber
            \end{equation}
        \end{subequations}

        Given non-negative dual variables $\delta_{(\cdot)}^-, \delta_{(\cdot)}^+, \beta_{(\cdot)}^-, \beta_{(\cdot)}^+$, where $(\cdot)$ represents intervals $\{u,v,x,y,z\}$, the proof is straightforward.
\end{proof}

\begin{table}[!t]
        \centering
        \captionsetup{justification=centering, labelsep=period, font=footnotesize, textfont=sc}
\caption{\vspace{0mm}Storage Unit Control Policy as Price Taker in Two-Interval Bidding \vspace{0mm}}
        \begin{tabular}{p{0.27\columnwidth}p{0.11\columnwidth}p{0.11\columnwidth}p{0.11\columnwidth}p{0.11\columnwidth}}
            \toprule
            \multirow{2}{*}{Scenario} & \multicolumn{2}{l}{Interval $1$} & \multicolumn{2}{l}{Interval $2$}  \\ \cmidrule(r){2-3}\cmidrule(r){4-5}
                                     & $p_1^*$            & $b_1^*$            & $p_2^*$            & $b_2^*$                             \\ \midrule
             $\hat{\lambda}_1 >  \frac{\hat{\lambda}_2}{\eta^2}$ & $\bar{P}\eta^2$ & 0 & 0& $\bar{P}$\\
             $  \hat{\lambda}_2 \eta^2\le \hat{\lambda}_1 \le \frac{\hat{\lambda}_2}{\eta^2}$ & 0 & 0 & 0& 0\\
            $\hat{\lambda}_1 <  \hat{\lambda}_2 \eta^2$ & 0 & $\bar{P}$ & $\bar{P}\eta^2$&0 \\
            \bottomrule
        \end{tabular}
        \label{table:taker}
        \vspace{-4mm}
\end{table}

\begin{table*}[!t]
\centering
\captionsetup{justification=centering, labelsep=period, font=footnotesize, textfont=sc}
\caption{\vspace{0mm}Storage Unit Control Policy as Price Maker in Two-Interval Bidding \vspace{0mm}}
\footnotesize
\begin{tabular}{p{0.13\textwidth}p{0.22\textwidth}p{0.12\textwidth}p{0.12\textwidth}p{0.12\textwidth}p{0.12\textwidth}}
        \toprule
          \multicolumn{2}{l}{\multirow{2}{*}{Scenario}} & \multicolumn{2}{l}{Interval $1$} & \multicolumn{2}{l}{Interval $2$}  \\ \cmidrule(r){3-4}\cmidrule(r){5-6}
               &                 & $p_1^*$            & $b_1^*$            & $p_2^*$            & $b_2^*$                                \\ \midrule
        
        $  \bar{\lambda}_1>\frac{\bar{\lambda}_2}{\eta^2} $ &$\bar{\lambda}_1  - 2 \alpha_1\bar{P} \eta^2 \ge  \frac{\bar{\lambda}_2 + 2 \alpha_2 \bar{P}}{\eta^2} $ & $\bar{P}\eta^2$ & 0 & 0& $\bar{P}$\\
         &$\bar{\lambda}_1  - 2 \alpha_1\bar{P}\eta^2 <  \frac{\bar{\lambda}_2 + 2 \alpha_2 \bar{P}}{\eta^2} $  & $\frac{\bar{\lambda}_1 -  \frac{\bar{\lambda}_2 }{\eta^2}}{2(\alpha_1  +\frac{\alpha_2}{\eta^4})}$ & 0 & 0& $\frac{\bar{\lambda}_1 -  \frac{\bar{\lambda}_2 }{\eta^2}}{2(\alpha_1  +\frac{\alpha_2}{\eta^4})\eta^2}$ \\ \midrule
         $\bar{\lambda}_2 \eta^2\le \bar{\lambda}_1 \le \frac{\bar{\lambda}_2}{\eta^2}$& & 0 & 0 & 0&0 \\ \midrule
         $\bar{\lambda}_1 < \bar{\lambda}_2\eta^2$ &$\frac{\bar{\lambda}_1 + 2\alpha_1 \bar{P}}{\eta^2} >  \bar{\lambda}_2 - 2\alpha_2\bar{P}\eta^2$  & 0 & $\frac{-\bar{\lambda}_1 +  \bar{\lambda}_2 \eta^2}{2(\alpha_1  +\alpha_2\eta^4)}$ & $\frac{(-\bar{\lambda}_1 +  \bar{\lambda}_2 \eta^2)\eta^2}{2(\alpha_1  +\alpha_2\eta^4)}$&0 \\
         &$\frac{\bar{\lambda}_1 + 2\alpha_1 \bar{P}}{\eta^2} \le  \bar{\lambda}_2 - 2\alpha_2\bar{P}\eta^2$ & 0 & $\bar{P}$ & $\bar{P}\eta^2$&0 \\
        \bottomrule
\end{tabular}
\label{table:maker}
\vspace{-3mm}
\end{table*}

We illustrate the bidding strategies of price takers based on model~\eqref{eq:sses3taker} through the two-interval bidding scenario, with the control policy detailed in Table~\ref{table:taker}. The decision-making process is primarily influenced by the price difference between the two intervals. Considering the energy loss during the charging/discharging cycle, the storage unit operation is justified only if the losses are offset by the profits derived from arbitraging the price difference. This arbitrage strategy indicates that the storage unit will fully utilize its capacity for profit as long as it remains economically viable, without engaging in unjustified withholding, i.e., no partial operation below $\bar{P}\eta^2$.

\subsection{Withholding as Price Maker}
We now explore the scenario in which the storage unit operates as a price maker, anticipating the effects of its bidding behavior on the market clearing price, using the following bidding model:
\begin{align}
\label{eq:sses2maker}
        \underset{p_t,b_t}{\text{maximize}}&\quad \sum_{t \in \mathcal{T}}(\bar{\lambda}_t - \alpha_t (p_t-b_t))(p_t-b_t)  \\
        \mathrm{s.t.}
        & \quad \eqref{eq:power_bal} - \eqref{eq:b_bar}. \nonumber                    
\end{align}

{Incorporating the impacts of market power on market clearing prices into the price maker's bidding model introduces quadratic terms into the objective function~\eqref{eq:sses2maker}, in contrast to the linear terms found in the price-taker's model as~\eqref{eq:obj_profit_taker}. However, both problems are still convex.}

Thus, the corresponding KKT conditions are ($\forall t \in \mathcal{T}$):
\begin{subequations}
  \begin{align}
    & \frac{\partial L}{\partial p_{t}} = \bar{\lambda}_{t} -2\alpha_t p_t+2\alpha_tb_t +\frac{\theta}{\eta}+ \delta_{t}^- - \delta_{t}^+ = 0\label{eq:kkt1_p}\\
    & \frac{\partial L}{\partial b_{t}} = -\bar{\lambda}_{t} +2\alpha_tp_t-2\alpha_t b_t-\theta\eta+ \beta_{t}^- - \beta_{t}^+ = 0\label{eq:kkt1_b}\\
    & \eqref{eq:kkt_bal} - \eqref{eq:kkt_dual} \nonumber         
  \end{align}
  \label{eq:kktmaker-3}
\end{subequations}
where Eqs.\eqref{eq:kkt1_p} and \eqref{eq:kkt1_b} correspond to stationarity conditions, and the rest are the same as that of the optimality conditions in~\eqref{eq:kkt_taker}.

\begin{proposition}
    For a strategic price maker making bidding decisions based on model~\eqref{eq:sses2maker}, throughout the period $\mathcal{T}$  
    \begin{enumerate}
        \item the bidding decisions $\{p^*_t,~b^*_t ,\, \forall t \in \mathcal{T}\}$ should include at least one interval $t$ at partial capacity, $0< p^*_t < \bar{P}$ or $0< b^*_t < \bar{P}$, except for the idle scenarios,
        \label{prop:4_1}
        \item there might exist multiple full or partial intervals.
        \label{prop:4_2}
\end{enumerate}
        \label{prop:partial_maker}
\end{proposition}
\begin{proof}
    \ref{prop:4_1}) follows similar arguments to Proposition~\ref{prop:partial}; \ref{prop:4_2}) is apparent following the proof for Proposition~\ref{prop:partial} based on the KKT conditions~\eqref{eq:kktmaker-3}.
\end{proof}

The two-interval bidding scenario for price makers based on model~\eqref{eq:sses2maker} is summarized in Table~\ref{table:maker}. 
Compared to the strategies of price takers as in Table~\ref{table:taker}, the control decisions for each non-idle scenario further splits into two sub-scenarios, either $ \bar{\lambda}_1>\frac{\bar{\lambda}_2}{\eta^2} $ or $\bar{\lambda}_1 < \bar{\lambda}_2\eta^2$.
Such decisions are established accounting for the economic losses associated with both charging/discharging energy loss and their potential impact on market prices, including both immediate and future contrast effects. Take the scenario where $ \bar{\lambda}_1>\frac{\bar{\lambda}_2}{\eta^2} $  for example. In addition to comparing prices based on charging and discharging efficiencies, i.e., $ \bar{\lambda}_1>\frac{\bar{\lambda}_2}{\eta^2} $, price makers also consider the marginal revenue from operational intervals. If the projected profit be substantial, i.e., with high marginal revenue as $\bar{\lambda}_1  - 2 \alpha_1\bar{P} \eta^2 \ge  \frac{\bar{\lambda}_2 + 2 \alpha_2 \bar{P}}{\eta^2} $ , the strategy aligns with that of price takers as operating at full capacity, suggesting that a larger price difference leads to greater profitability, sufficient even for arbitrage by price takers. Moreover, price makers, facing potentially lower marginal revenues, i.e., $\bar{\lambda}_1  - 2 \alpha_1\bar{P}\eta^2 <  \frac{\bar{\lambda}_2 + 2 \alpha_2 \bar{P}}{\eta^2} $, still possess the strategic flexibility to optimize control decisions for maximizing profit, exploiting their influence on market outcome.

\subsection{Three-Interval Bidding Scenario}

We further examine the three-interval bidding scenario to help illustrate the bidding pattern of price makers.
{Compared to the two-interval bidding discussed earlier, the three-interval scenario demonstrates how energy storage units strategically allocate their limited resources into a portfolio that maximizes total profit based on a series of prices over the scheduling period. The three-interval bidding establishes a foundation for analyzing longer-term bidding scenarios.}

Define the number of full capacity intervals throughout the scheduling period $\mathcal{T}$: $o=\sum_{t\in {\mathcal{T}}}\mathds{1}_{\{p_t=\bar{P}\}}+ \sum_{t\in{ \mathcal{T}}}\mathds{1}_{\{b_t=\bar{P}\}}$, and correspondingly for partial intervals: $s=\sum_{t\in {\mathcal{T}}}\mathds{1}_{\{0<p_t<\bar{P}\}}+ \sum_{t\in{ \mathcal{T}}}\mathds{1}_{\{0<b_t<\bar{P}\}}$.
\begin{corollary}
    For a strategic price maker making bidding decisions based on model~\eqref{eq:sses2maker}, consider the bidding case consisting three intervals, then if there exist two intervals $j_1,~j_2$ operating with partial capacity, $s=2$, and one full $t_0$, $o=1$ and $\bar{\lambda}_{j_2} > \bar{\lambda}_{j_1}, \bar{\lambda}_{j_2} > \bar{\lambda}_{t_0}$, a sufficient condition for this scenario is $\frac{\bar{\lambda}_{j_1}+2\alpha_{j_1}\bar{P}}{\eta^2}>\bar{\lambda}_{j_2}-2\alpha_{j_2}\bar{P}$.
    \label{cor:three-interval}
\end{corollary}
\begin{proof}
Note that charging and discharging intervals appear in pairs, then based on KKT conditions~\eqref{eq:kkt_com1}, \eqref{eq:kkt_com2}, \eqref{eq:kkt1_p} and \eqref{eq:kkt1_b} we have
\begin{subnumcases} {}
        -\bar{\lambda}_{t_0}-2\alpha_{t_0}\bar{P}-\theta\eta -\beta_{t_0}^+ =  0 \ (\mathrm{full}),\\
        -\bar{\lambda}_{j_1}-2\alpha_{j_1}b_{j_1}-\theta\eta =  0 \ (\mathrm{partial}) ,\label{eq:maker2}\\
        \bar{\lambda}_{j_2}-2\alpha_{j_2}p_{j_2}+\frac{\theta}{\eta}  =  0 \ (\mathrm{partial}).\label{eq:maker3}
\end{subnumcases}
        
Combining~\eqref{eq:maker2} and \eqref{eq:maker3}, we have
\begin{align}
        \frac{\bar{\lambda}_{j_1}+2\alpha_{j_1}b^*_{j_1}}{\eta^2}=\bar{\lambda}_{j_2}-2\alpha_{j_2}p^*_{j_2}. \label{eq:proof1}
\end{align}

Note that intervals $j_1,j_2$ are partial, $b^*_{j_1},p^*_{j_2}<\bar{P}$. Equation~\eqref{eq:proof1} becomes
\begin{align}
        \frac{\bar{\lambda}_{j_1}+2\alpha_{j_1}\bar{P}}{\eta^2}>\bar{\lambda}_{j_2}-2\alpha_{j_2}\bar{P}. \label{eq:proof2}
\end{align}

\end{proof}

For three-interval bidding, the inequality constraint~\eqref{eq:proof2} provides a sufficient condition for the scenario where $s=2,~ o=1$ when $\bar{\lambda}_{j_2} > \bar{\lambda}_{j_1},~ \bar{\lambda}_{j_2} > \bar{\lambda}_{t_0}$. It also implies  when the ultimate price difference is larger between intervals $j_1$ and $j_2$, there can be no less full intervals than partial. Such analysis can be easily extended to other scenarios. 
\begin{remark}
        Higher price difference, resulting in higher profitability, complicate the distinction between withholding behavior due to strategic arbitrage versus market power exercise. Both strategies exploit market inefficiencies, however, high profit margin can mask the source of earnings, whether derived from arbitrage or price manipulation.
        \label{remark:high_price_diff}
\end{remark}
        %
        

\subsection{Understanding Multi-Interval Bidding}

\begin{table*}[t]
        \centering
        \captionsetup{justification=centering, labelsep=period, font=footnotesize, textfont=sc}
        \caption{\vspace{0mm}Scenarios Price Maker Exercising Market Power by Conducting One-Interval Capacity Withholding \vspace{0mm}}
        \begin{tabular}{
                p{0.22\columnwidth}p{0.09\columnwidth}|
                p{0.11\columnwidth}p{0.11\columnwidth}p{0.11\columnwidth}p{0.11\columnwidth}p{0.11\columnwidth}|
                p{0.11\columnwidth}p{0.11\columnwidth}p{0.11\columnwidth}p{0.11\columnwidth}p{0.11\columnwidth}}
        \toprule
        \multicolumn{2}{l|}{\multirow{2}{*}{Scenario}} & \multicolumn{5}{l|}{Taker}                            & \multicolumn{5}{l}{Maker}                       \\ \cmidrule(r){3-7} \cmidrule(r){7-11} 
\multicolumn{2}{l|}{}                          & $a^{\dagger}$       & $b$       & $c$   & $d$       & $e$       & $a$       & $b$   & $c$ & $d$       & $e$       \\ \midrule 
$\hat{b}$ downgrade &    $p_t$ & $\bar{P}$ & \cellcolor{gray!25} $\bar{P}$ &       &           &           & $\bar{P}$ & \cellcolor{gray!25}$\hat{p}^{\ddagger}$     &           &           &           \\
                &    $b_t$ &           &           & \cellcolor{gray!25}$\hat{b}^{\ddagger}$ & $\bar{P}$ & $\bar{P}$ &           &           & \cellcolor{gray!25}$0$       & $\bar{P}$ & $\bar{P}$ \\ \midrule
$\hat{p}$ downgrade &    $p_t$ & $\bar{P}$ & $\bar{P}$ & \cellcolor{gray!25}$\hat{p}$ &           &           & $\bar{P}$ & $\bar{P}$ & \cellcolor{gray!25}$0$       &           &           \\
                &    $b_t$ &           &           &       & \cellcolor{gray!25}$\bar{P}$ & $\bar{P}$ &           &           &           & \cellcolor{gray!25}$\hat{b}$     & $\bar{P}$ \\\midrule 
$\hat{b}$ upgrade   & $p_t$ & $\bar{P}$ & $\bar{P}$ &       &           &           & $\bar{P}$ & $\bar{P}$ &           &           &           \\
                &    $b_t$ &           &           & \cellcolor{gray!25}$\hat{b}$ & \cellcolor{gray!25}$\hat{b}$     & $\bar{P}$ &           &           & \cellcolor{gray!25}$\hat{b}$     & \cellcolor{gray!25}$\bar{P}$ & $\bar{P}$ \\ 
                \midrule 
$\hat{p}$ upgrade   &   $p_t$ & $\bar{P}$ &\cellcolor{gray!25} $\hat{p}$     & \cellcolor{gray!25}$\hat{p}$ &           &           & $\bar{P}$ & \cellcolor{gray!25}$\bar{P}$ & \cellcolor{gray!25}$\hat{p}$     &           &           \\
                &    $b_t$ &           &           &       & $\bar{P}$ & $\bar{P}$ &           &           &           & $\bar{P}$ & $\bar{P}$ \\
                \midrule
$\hat{b}$ switch    &  $p_t$ &           &           &       &           &           & $\bar{P}$ & $\bar{P}$ & $0$       &           &           \\
                &    $b_t$ &           &           &       &           &           &           &           & $0$       & \cellcolor{gray!25}$\bar{P}$ &\cellcolor{gray!25} $\hat{b}$     \\\cmidrule(r){1-2} \cmidrule(r){8-12} 
                &  $p_t$ & $\bar{P}$ & $\bar{P}$ & $0$   &           &           & $\bar{P}$ & $\bar{P}$ & $0$       &           &           \\
                &    $b_t$ &           &           & \cellcolor{gray!25}$0$   & \cellcolor{gray!25}$\hat{b}$     & \cellcolor{gray!25}$\bar{P}$ &                      &           &\cellcolor{gray!25}$\hat{b}$     & \cellcolor{gray!25}$0$          & $\bar{P}$ \\ \cmidrule(r){1-2} \cmidrule(r){8-12} 
                &  $p_t$ &           &           &       &           &           & $\bar{P}$ & $\bar{P}$ & $0$       &           &           \\
                &    $b_t$ &           &           &       &           &           &           &           & \cellcolor{gray!25}$\bar{P}$ & $\hat{b}$     & \cellcolor{gray!25}$0$       \\\midrule 

                $\hat{p}$ switch    &  $p_t$ &           &           &       &           &           &\cellcolor{gray!25} $\hat{p}$     & \cellcolor{gray!25}$\bar{P}$ & $0$       &           &           \\
                &    $b_t$ &           &           &       &           &           &           &           & $0$       & $\bar{P}$ & $\bar{P}$ \\\cmidrule(r){1-2} \cmidrule(r){8-12} 
                &  $p_t$ & \cellcolor{gray!25}$\bar{P}$ & \cellcolor{gray!25}$\hat{p}$     & \cellcolor{gray!25}$0$    &           &           & $\bar{P}$ & \cellcolor{gray!25}$0$          & \cellcolor{gray!25}$\hat{p}$     &           &           \\
                &    $b_t$ &           &           & $0$   & $\bar{P}$ & $\bar{P}$ &           &           & $0$       & $\bar{P}$ & $\bar{P}$ \\\cmidrule(r){1-2} \cmidrule(r){8-12} 

                &  $p_t$ &           &           &       &           &           & \cellcolor{gray!25}$0$       & $\hat{p}$     &\cellcolor{gray!25} $\bar{P}$ &           &           \\
                &    $b_t$ &           &           &       &           &           &           &           & $0$       & $\bar{P}$ & $\bar{P}$\\ \bottomrule
                \multicolumn{12}{l}{\color{black}$^{\dagger}$ $\{a,b,c,d,e\}$ represent different random intervals.}\\
                \multicolumn{12}{l}{\color{black}$^{\ddagger}$ $\hat{p}$ and $\hat{b}$ give the generic notation for the withheld power output, i.e., $\hat{p},\, \hat{b}\in (0,\bar{P})$.}
        \end{tabular}

        \label{table:counterex}
\end{table*}

The  \textit{ex-post} analysis is established in two phases. {First, 
the market operator counts and compares the number of full or partial intervals, 
as illustrated in Fig.\ref{fig:market_role} summarizing the discussion from the previous subsections.} For price takers, as suggested in Propositions~\ref{prop:full} and~\ref{prop:partial}, they typically participate in the market with the profiles where the number of full intervals are no more the that of partial ones, including the idle scenarios. On the other hand, for price makers, Proposition~\ref{prop:partial_maker} and Corollary~\ref{cor:three-interval} indicate that most scenarios feature the number of partial intervals no less than that of full ones, suggesting the exercise of market power through capacity withholding.  Exceptions occur where the number of full intervals exceed that of partial ones, notably in situations with significant price differentials, as detailed in Remark~\ref{remark:high_price_diff}. 
{Secondly, examining price-decision relationship across the intervals reveals the bidding strategy adopted by the market participant, indicating whether it involves the exercise of market power.}
With this analysis in place, we revisit Theorem~\ref{theorem:oneinterval} and present its proof as follows:

\begin{figure}[!t]
        \centering
        \includegraphics[width=0.45\textwidth]{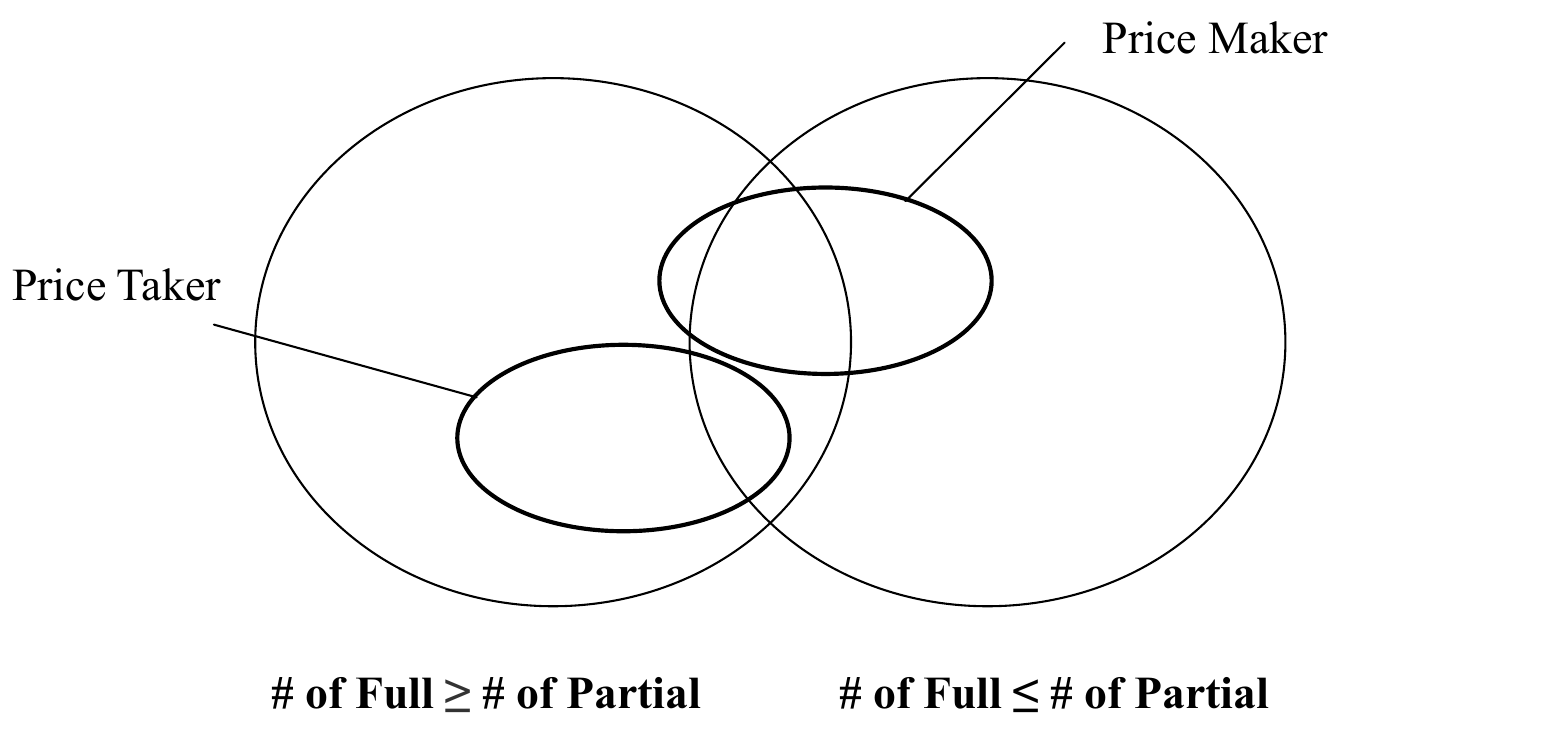}
        \caption{\small Bidding pattern regarding capacity withholding as a price taker or a price maker given strictly heterogeneous prices.}
        \label{fig:market_role}
        \vspace{-3mm}
\end{figure}

\begin{proof}[Theorem~\ref{theorem:oneinterval}]
We first discuss the scenarios for price takers and price makers within the framework of Assumptions~\ref{ass:short_period}--\ref{ass:nonneg_price}, then explore more practical cases following the relaxation of these assumptions.

For price takers, from Proposition~\ref{prop:partial} we've covered that given strictly heterogeneous prices, there exists only one partial interval, which establishes a sufficient condition for Theorem~\ref{theorem:oneinterval}. Relaxing this constraint to allow for instances of identical prices potentially leads to the presence of additional partial intervals within the scheduling period. Therefore, Theorem~\ref{theorem:oneinterval} is applicable to price takers given these assumptions. 

For price makers, Proposition~\ref{prop:partial_maker} states that the number of partial intervals is always no less than one, i.e., $s\ge 1$. {When $s=1$, assuming small price sensitivity parameter $\alpha_t$, we can conclude that the bidding profiles from price makers are identical to those observed among price takers, thus the price maker is not evidently exercising market power. However, scenarios exist where, despite the profile showing only one withholding interval, the price maker \emph{is} exercising market power as summarized in Table~\ref{table:counterex}.
We take the scenario \emph{$\hat{b}$ downgrade} from Table~\ref{table:counterex} for example to explain how Theorem~\ref{theorem:oneinterval} can identify the one-interval-withholding market power exercise:

For the price taker, in order to derive the bidding decisions $p_b^* = \bar{P}, \,~ b_c^* =\hat{b}$, the corresponding clearing prices need to satisfy $\hat{\lambda}_b>\frac{\hat{\lambda}_c }{\eta^2}$ as illustrated in Proposition~\ref{prop:taker_price}. Similarly, for the price maker, following the KKT conditions~\eqref{eq:kktmaker-3}, the bidding decisions $p_b^* = \hat{p}, \,~ b_c^* = 0$ are derived under the condition:
        \begin{align}
                \hat{\lambda}_c - \alpha_c < \hat{\lambda}_b + \alpha_b (\bar{P}- 2 \hat{p})< \frac{\hat{\lambda}_c - \alpha_c  }{\eta^2}, 
        \end{align}
         and the resulting clearing prices are:
        \begin{align}
                & \hat{\lambda}'_b  = \hat{\lambda}_b + \alpha_b (\bar{P}- \hat{p}),\nonumber \\
                &\hat{\lambda}'_c  = \hat{\lambda}_c - \alpha_c  \hat{b}. \nonumber 
        \end{align}
        Given $\alpha_b(\bar{P}- \hat{p})>0$ and $\alpha_c \hat{b}>0$, we have $\hat{\lambda}'_b>\frac{\hat{\lambda}_c' }{\eta^2}$. Thus, the price-decision relationship between $(\hat{\lambda}'_b,\hat{\lambda}'_c)$ and $(p_b^* = \hat{p}, b_c^* = 0)$ violates Proposition~\ref{prop:taker_price}. The proof for the remaining  scenarios follow similar arguments.
        
}

Therefore, Theorem~\ref{theorem:oneinterval} holds for price makers under the given Assumptions~\ref{ass:short_period}--\ref{ass:nonneg_price}.
Specially, if the unit is held idle, no market power is exercised.

Next, we generalize the conditions following the relaxation of the initial assumptions. Dropping Assumption~\ref{ass:short_period} generates scenarios where energy storage unit output is limited by its energy capacity, potentially leading to the occurrence of additional partial blocks. This modification does not compromise the validity of Theorem~\ref{theorem:oneinterval}.
Regarding Assumption~\ref{ass:storage_idle}, incorporating scenarios of active operation during zero-profit periods allows for flexible dispatch decisions $p^*_t, b^*_t \in [0,\bar{P}]$, which also does not alter the conclusions drawn in Theorem~\ref{theorem:oneinterval}. Assumption~\ref{ass:nonneg_price} is initially established to simplify the analysis of KKT optimality conditions. Negative prices enforce storage unit no discharging according to~\eqref{eq:storage_idle}, therefore $p^*_t = 0$ for certain intervals. The removal of such assumption does not impact the conclusions reached in prior subsections. We defer further discussion of this topic to readers with an interest in exploring it in greater detail.
 \end{proof}

\section{NUMERICAL EXPERIMENTS}
We validate the proposed framework for energy storage unit bidding by simulating a 24-interval dispatch in the day-ahead market. Our simulation features a 2.5MW/10MWh storage unit that starts and ends at 50\% SoC, with charging and discharging efficiencies set at 0.9. Note that the energy capacity constraint~\eqref{eq:soc_bar} is taken into account in the simulation.
We compare the bidding profiles and resulting profits between price takers and price makers, using price data from a winter day in New York City in 2018, as reported by NYISO~\cite{NYISO_price}. This price series serves as the benchmark for competitive market clearing prices. 
Nominal clearing prices $\bar{\lambda}_t$ are calculated by assuming that non-withholding power supply from the price taker is contributed to the market, thereby leading to  competitive prices.
The price sensitivity parameter $\alpha_t$ is modeled to be linearly proportional to the competitive price level, reflecting the load level and the slope of the remaining supply function, with an average value of \$1.00/MWh/MWh for low market power price makers and \$2.00/MWh/MWh for high market power cases.
Note that this parameter is amplified for the purposes of validating our framework in the simulation and would be expected to be significantly lower in practical scenarios~\cite{reinisch2006}. 

\begin{figure}[t]
        \begin{subfigure}[t]{\columnwidth}
            \centering    
            \includegraphics[width=\columnwidth]{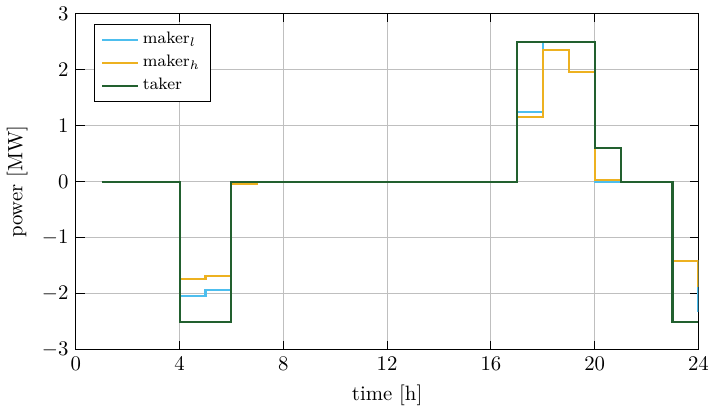}
            \vspace{-5mm}
            \caption{\small  Storage power output\vspace{-0mm}}
            \label{fig:power}
        \end{subfigure}
        \vfill
         \begin{subfigure}[t]{\columnwidth}
            \centering    
            \vspace{1mm}
            \includegraphics[width=0.97\columnwidth]{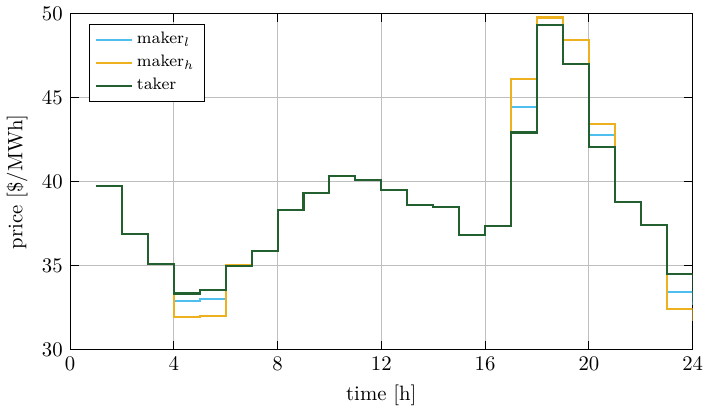}
            \caption{\small Market clearing price\vspace{-0mm}}
            \label{fig:price}
        \end{subfigure}
        \caption{\small Storage unit control policy and the resulting market clearing price considering different market participants as price maker with low market power ($\mathrm{maker}_l$), high market power ($\mathrm{maker}_h$), and price taker: (a) storage power output (positive values mean that unit is discharging), (b) market clearing price. \vspace{-1mm}}
        \label{fig:sim}
\end{figure}

The market clearing results are presented in Fig.~\ref{fig:sim}. Figure~\ref{fig:sim}(\subref{fig:power}) illustrates the capacity withholding behavior of price makers with both low and high market power in comparison to price takers. Figure~\ref{fig:sim}(\subref{fig:price}) displays the resulting shifts in market clearing prices from the competitive benchmark, highlighting the exercise of market power. 

Figure~\ref{fig:sim}(\subref{fig:power}) illustrates that, as a price maker, capacity withholding occurs mainly during periods of peak or valley prices to arbitrage price differences. For example, during peak hours, price makers may withhold storage discharging to raise prices, e.g., 5-9pm, as reflected by the corresponding market clearing prices in Fig.~\ref{fig:sim}(\subref{fig:price}). Conversely, during off-peak hours, they might withhold charging to lower prices. Furthermore, a price maker with a higher degree of market power can exert a more significant influence on the market. This is demonstrated by the more aggressive withholding behavior and the subsequent alterations in clearing prices during both peak and off-peak periods. Notably, even when applying the price maker's bidding model~\eqref{eq:sses2maker}, there are instances where the price maker bids similarly to the price taker, leading to identical power outputs and market clearing prices, e.g., 8am-5pm. This indicates that they are not actively exercising market power during these times. For the price taker, there is only one interval of withholding, i.e., 8-9pm, consistent with Theorem~\ref{theorem:oneinterval}. This theorem provides a criterion for identifying evident instances of market power exercise. Specially, in some cases, such as from 6-7pm,  the power production from the price maker might exceed that of the price taker, which could seem counterintuitive at first glance but could be interpreted as strategic practices aimed at maximizing  profit. The profits generated under these scenarios are summarized in Table~\ref{table:profit}. 

\begin{table}[!t]
        \centering
        \captionsetup{justification=centering, labelsep=period, font=footnotesize, textfont=sc}
\caption{\vspace{0mm} Profit of Market Participants under Different Level of Market Power  \vspace{0mm}}
        \begin{tabular}{p{0.4\columnwidth}p{0.22\columnwidth}p{0.22\columnwidth}}
            \toprule
            Scenario&Price Taker ($\$$)                           & Price Maker ($\$$)                  \\ \midrule
            No market power &37.95 & -- \\
            Low market power & 47.50  & 42.02 \\
            High market power & 66.66& 49.11\\
            \bottomrule
        \end{tabular}
        \label{table:profit}
        \vspace{-3mm}
\end{table}

Table~\ref{table:profit} indicates that the price maker exercises market power to gain additional profit. When possessing a higher level of market power, the price maker's earnings are 76\% greater than in scenarios where they act as a price taker (\$49.11 vs \$37.95), and 25\% greater with a lower level of market power (\$42.02 vs \$37.95). Interestingly, in markets where a price maker is present, price takers may achieve higher profits than the price maker (\$42.02 vs \$47.50 and \$49.11 vs \$66.66), aligning with findings discussed in~\cite{wu2023}.
This observation highlights the vulnerability of market inefficiencies to strategic exploitation.

\section{CONCLUSIONS}
We examine the multi-interval strategic bidding withholding of energy storage units adopting a self-scheduling model, considering both price takers and price makers. For price makers, we introduce a bidding strategy that anticipates market prices through a price sensitivity analysis. The proposed framework serves as an \emph{ex-post} market monitoring tool, allowing market operators to distinguish market power exercise from competitive withholding by observing the withholding instances within the bidding profiles and the corresponding clearing prices.  Our findings support the economic bidding behaviors commonly seen among storage units and reveal that significant price fluctuations present substantial profit opportunities for both market power holders and strategic profit-seekers.
In future work, we aim to enhance our model by incorporating uncertainty in price forecasts into the framework and develop corresponding criteria for market power assessment.

\section{Acknowledgements}

{JA acknowledges funding from NSF grants ECCS 2144634 and ECCS 2047213 as well as the Columbia Data Science Institute. Bolun Xu acknowledges funding from NSF grants ECCS 2239046.}


\bibliographystyle{IEEEtran}
\bibliography{market.bib}

\begin{thebibliography}{10}
\providecommand{\url}[1]{#1}
\csname url@rmstyle\endcsname
\providecommand{\newblock}{\relax}
\providecommand{\bibinfo}[2]{#2}
\providecommand\BIBentrySTDinterwordspacing{\spaceskip=0pt\relax}
\providecommand\BIBentryALTinterwordstretchfactor{4}
\providecommand\BIBentryALTinterwordspacing{\spaceskip=\fontdimen2\font plus
\BIBentryALTinterwordstretchfactor\fontdimen3\font minus \fontdimen4\font\relax}
\providecommand\BIBforeignlanguage[2]{{%
\expandafter\ifx\csname l@#1\endcsname\relax
\typeout{** WARNING: IEEEtran.bst: No hyphenation pattern has been}%
\typeout{** loaded for the language `#1'. Using the pattern for}%
\typeout{** the default language instead.}%
\else
\language=\csname l@#1\endcsname
\fi
#2}}

\bibitem{sakti2018review}
A.~Sakti, A.~Botterud, and F.~O’Sullivan, ``Review of wholesale markets and regulations for advanced energy storage services in the united states: Current status and path forward,'' \emph{Energy policy}, vol. 120, pp. 569--579, 2018.

\bibitem{caiso_es}
\BIBentryALTinterwordspacing
{California ISO }, ``Special report on energy storage,'' 2023. [Online]. Available: \url{https://www.caiso.com/Documents/2022-Special-Report-on-Battery-Storage-Jul-7-2023.pdf}
\BIBentrySTDinterwordspacing

\bibitem{eia}
\BIBentryALTinterwordspacing
{U.S. Energy Information Administration}, ``Battery systems on the u.s. power grid are increasingly used to respond to price,'' 2022. [Online]. Available: \url{https://www.eia.gov/todayinenergy/detail.php?id=53199}
\BIBentrySTDinterwordspacing

\bibitem{ma2021data}
W.~Ma and B.~Xu, ``A data-driven nonlinear recharge controller for energy storage in frequency regulation,'' in \emph{2021 IEEE Power \& Energy Society General Meeting (PESGM)}, 2021, pp. 1--5.

\bibitem{mohsenian2015coordinated}
H.~Mohsenian-Rad, ``Coordinated price-maker operation of large energy storage units in nodal energy markets,'' \emph{IEEE Transactions on Power Systems}, vol.~31, no.~1, pp. 786--797, 2015.

\bibitem{baker2023transferable}
Y.~Baker, N.~Zheng, and B.~Xu, ``Transferable energy storage bidder,'' \emph{IEEE Transactions on Power Systems}, 2023.

\bibitem{chen2021pricing}
C.~Chen, L.~Tong, and Y.~Guo, ``Pricing energy storage in real-time market,'' in \emph{2021 IEEE Power \& Energy Society General Meeting (PESGM)}, 2021, pp. 1--5.

\bibitem{krishnamurthy2017energy}
D.~Krishnamurthy, C.~Uckun, Z.~Zhou, P.~R. Thimmapuram, and A.~Botterud, ``Energy storage arbitrage under day-ahead and real-time price uncertainty,'' \emph{IEEE Transactions on Power Systems}, vol.~33, no.~1, pp. 84--93, 2017.

\bibitem{bansal2022market}
R.~K. Bansal, P.~You, D.~F. Gayme, and E.~Mallada, ``A market mechanism for truthful bidding with energy storage,'' \emph{Electric Power Systems Research}, vol. 211, p. 108284, 2022.

\bibitem{baldick2005design}
R.~Baldick, U.~Helman, B.~F. Hobbs, and R.~P. O'Neill, ``Design of efficient generation markets,'' \emph{Proceedings of the IEEE}, vol.~93, no.~11, pp. 1998--2012, 2005.

\bibitem{harvey2001market}
S.~M. Harvey and W.~W. Hogan, ``Market power and withholding,'' \emph{Harvard University, Cambridge, MA}, 2001.

\bibitem{reinisch2006}
W.~Reinisch and T.~Tezuka, ``Market power and trading strategies on the electricity market: a market design view,'' \emph{IEEE Transactions on Power Systems}, vol.~21, no.~3, pp. 1180--1190, 2006.

\bibitem{baldick2004}
R.~Baldick, R.~Grant, and E.~Kahn, ``Theory and application of linear supply function equilibrium in electricity markets,'' \emph{Journal of Regulatory Economics}, vol.~25, pp. 143--167, 2004.

\bibitem{xu2020}
B.~Xu, M.~Korpås, and A.~Botterud, ``Operational valuation of energy storage under multi-stage price uncertainties,'' in \emph{2020 59th IEEE Conference on Decision and Control (CDC)}, 2020, pp. 55--60.

\bibitem{NYISO_price}
\BIBentryALTinterwordspacing
{New York ISO}, ``Energy market \& operational data,'' 2024. [Online]. Available: \url{https://www.nyiso.com/energy-market-operational-data}
\BIBentrySTDinterwordspacing

\bibitem{wu2023}
Y.~Wu, J.~Kim, and J.~Anderson, ``Mitigation-aware bidding strategies in electricity markets,'' in \emph{2023 IEEE Power \& Energy Society General Meeting (PESGM)}, 2023, pp. 1--5.

\end{thebibliography}

\end{document}